\documentclass[twocolumn]{aastex7}
\hypersetup{linkcolor=red,citecolor=blue,filecolor=cyan,urlcolor=magenta}
\usepackage{nameref}
\usepackage{amsmath}
\usepackage{placeins}
\usepackage{booktabs} 
\usepackage{siunitx} 
\long\def\symbolfootnote[#1]#2{\begingroup%
\def\thefootnote{\fnsymbol{footnote}}\footnote[#1]{#2}\endgroup}
\begin{document}

\newcommand{\vdag}{(v)^\dagger}
\newcommand\aastex{AAS\TeX}
\newcommand\latex{La\TeX}
\newcommand{\msun}{M_{\odot}}
\newcommand{\mstar}{M_{\star}}
  
\definecolor{ForestGreen}{RGB}{34,139,34}
\def\tw#1 {{\textcolor{ForestGreen}{#1}}\ }
\def\red#1 {{\textcolor{red}{#1}}\ }
\def\blue#1{{\textcolor{black}{#1}}\ }
\def\review2#1{{\textcolor{black}{#1}}\ }

\title{A census of quiescent galaxies across $0.5 < z < 8$ with JWST/MIRI: Mass-dependent number density evolution of quiescent galaxies in the early Universe}

\author[0009-0008-4971-035X, sname=Yang, gname=Tiancheng]{Tiancheng Yang}
\affiliation{School of Astronomy and Space Science, Nanjing University, Nanjing, Jiangsu 210093, China}
\affiliation{Key Laboratory of Modern Astronomy and Astrophysics (Nanjing University), Ministry of Education, Nanjing 210093, China}
\email{tcy@smail.nju.edu.cn}

\author[0000-0002-2504-2421,sname=Wang,gname=Tao]{Tao Wang}
\affiliation{School of Astronomy and Space Science, Nanjing University, Nanjing, Jiangsu 210093, China}
\affiliation{Key Laboratory of Modern Astronomy and Astrophysics (Nanjing University), Ministry of Education, Nanjing 210093, China}
\email[show]{taowang@nju.edu.cn}

\author{Ke Xu}
\affiliation{School of Astronomy and Space Science, Nanjing University, Nanjing, Jiangsu 210093, China}
\affiliation{Key Laboratory of Modern Astronomy and Astrophysics (Nanjing University), Ministry of Education, Nanjing 210093, China}
\email[]{kexu@smail.nju.edu.cn}

\author{Hanwen Sun}
\affiliation{School of Astronomy and Space Science, Nanjing University, Nanjing, Jiangsu 210093, China}
\affiliation{Key Laboratory of Modern Astronomy and Astrophysics (Nanjing University), Ministry of Education, Nanjing 210093, China}
\email[]{hanwensun@smail.nju.edu.cn}

\author{Luwenjia Zhou}
\affiliation{School of Astronomy and Space Science, Nanjing University, Nanjing, Jiangsu 210093, China}
\affiliation{Key Laboratory of Modern Astronomy and Astrophysics (Nanjing University), Ministry of Education, Nanjing 210093, China}
\email[]{wenjia@nju.edu.cn}

\author[0000-0003-3864-068X,sname=Xie, gname=Lizhi]{Lizhi Xie}
\affiliation{Tianjin Normal University, Binshuixidao 393, Xiqing, 300387, Tianjin, People's Republic of China;}
\email[]{xielizhi.1988@gmail.com}

\author{Gabriella De Lucia}
\affiliation{INAF—Astronomical Observatory of Trieste, via G.B. Tiepolo 11, I-34143 Trieste, Italy}
\affiliation{IFPU—Institute for Fundamental Physics of the Universe, via Beirut 2, 34151, Trieste, Italy}
\email[]{gabriella.delucia@inaf.it}

\author{Claudia del P. Lagos}
\affiliation{International Centre for Radio Astronomy Research (ICRAR), M468, University of Western Australia, 35 Stirling Hwy, Crawley, WA 6009, Australia}
\affiliation{ARC Centre of Excellence for All Sky Astrophysics in 3 Dimensions (ASTRO 3D)}
\email[]{claudia.lagos@uwa.edu.au}

\author[0000-0002-3775-0484,sname=Wang,gname=Kai]{Kai Wang}
\affiliation{Institute for Computational Cosmology, Department of Physics, Durham University, South Road, Durham, DH1 3LE, UK}
\affiliation{Centre for Extragalactic Astronomy, Department of Physics, Durham University, South Road, Durham DH1 3LE, UK}
\email[]{wkcosmology@gmail.com}

\author{Fabio Fontanot}
\affiliation{INAF—Astronomical Observatory of Trieste, via G.B. Tiepolo 11, I-34143 Trieste, Italy}
\affiliation{IFPU—Institute for Fundamental Physics of the Universe, via Beirut 2, 34151, Trieste, Italy}
\email[]{fabio.fontanot@inaf.it}

\author{Qi Guo}
\affiliation{Key Laboratory for Computational Astrophysics, National Astronomical Observatories, Chinese Academy of Sciences, Beijing 100012, People’s Republic of
China}
\affiliation{School of Astronomy and Space Sciences, University of Chinese Academy of Sciences, 19A Yuquan Road, Beijing 100049, People’s Republic of China}
\affiliation{Institute for Frontiers in Astronomy and Astrophysics, Beijing Normal University, Beijing 102206, People’s Republic of China}
\affiliation{School of Physics and Astronomy, Beijing Normal University, Beijing 100875, People’s Republic of China}
\email[]{guoqi@nao.cas.cn}

\author{Yuxuan Wu}
\affiliation{School of Astronomy and Space Science, Nanjing University, Nanjing, Jiangsu 210093, China}
\affiliation{Key Laboratory of Modern Astronomy and Astrophysics (Nanjing University), Ministry of Education, Nanjing 210093, China}
\email[]{yuxuan.wu@smail.nju.edu.cn}

\author{shiying Lu}
\affiliation{School of Mathematics and Physics, Anqing Normal University, Anqing 246133, China}
\affiliation{Institute of Astronomy and Astrophysics, Anqing Normal University, Anqing 246133, China}
\affiliation{Key Laboratory of Modern Astronomy and Astrophysics (Nanjing University), Ministry of Education, Nanjing 210093, China}
\email[]{976311399@qq.com}

\author{Longyue Chen}
\affiliation{School of Astronomy and Space Science, Nanjing University, Nanjing, Jiangsu 210093, China}
\affiliation{Key Laboratory of Modern Astronomy and Astrophysics (Nanjing University), Ministry of Education, Nanjing 210093, China}
 \email[]{652023260002@smail.nju.edu.cn}

 \author{Michaela Hirschmann}
 \affiliation{1Institute of Physics, Laboratory for Galaxy Evolution, EPFL, Observatoire de Sauverny, Chemin Pegasi 51, CH-1290 Versoix, Switzerland}
 \email[]{michaela.hirschmann@epfl.ch}


\begin{abstract}  

Recent JWST observations have revealed a large population of quiescent galaxies (QGs) at high redshift ($z \sim 4-8$), challenging current models of early galaxy formation and quenching. Accurate number density estimates are crucial but remain uncertain. We present a systematic study of QGs at $0.5 < z < 8$ using a mass-complete sample from the JWST/PRIMER survey with deep NIRCam and MIRI imaging. \blue{We demonstrate that MIRI photometry is {important} for refining the QG sample: it helps to mitigate contamination from dusty star-forming galaxies in the high-mass regime at $z \sim 3-5$ and aids in recovering lower-mass QG candidates at $z > 5$ that are often missed without including MIRI data.} We find that the evolution of the QG number density is strongly mass-dependent. The density of massive QGs ($\log (M_{\star}/M_{\odot}) > 10.6$) declines rapidly, falling from $n \approx 1.32\times10^{-5}~~\mathrm{Mpc^{-3}}$ at $z \sim 3-4$ to $n \sim 1 \times10^{-6}~~\mathrm{Mpc^{-3}}$ at $z \sim 6$, and becomes negligible at $z > 6$. In contrast, low-mass QGs ($9.5 < \log (M_{\star}/M_{\odot}) < 10.6$) exhibit a constant number density of $n \sim 2\times10^{-6}~\mathrm{Mpc^{-3}}$ across the redshift range $z = 4-8$. \blue{This plateau suggests that these high-redshift, low-mass QGs may be galaxies undergoing temporary quenching episodes, likely subject to rejuvenation upon future gas accretion.} Comparisons with leading galaxy formation models reveal significant tensions: most models underestimate the abundance of massive QGs at $z > 4$ and fail to reproduce the flat density evolution observed for the low-mass population.


\end{abstract}


\keywords{\uat{Galaxy evolution}{594} --- \uat{High-redshift galaxies}{734} --- \uat{Galaxy quenching}{2040} --- \uat{Post-starburst galaxies}{2176}}

\section{Introduction}
\label{sec:intro}

Understanding the origin of galaxy bimodality—specifically, the physical mechanisms driving the cessation of star formation—remains a fundamental challenge in extragalactic astronomy. Consequently, obtaining stringent constraints on the cosmic evolution of the number density and physical properties of quiescent galaxies (QGs) is essential. Over the past two decades, the evolution of the QG number density and quiescent fraction for galaxies with $\mstar \gtrsim 10^{10} \msun$ at $0 < z < 3$ has been well established \citep{muzzin_evolution_2013,straatman_fourstar_2016,brammer_number_2011}. Observations consistently show that the number density of QGs declines rapidly with increasing redshift, while the quiescent fraction increases with stellar mass up to $z \sim 3$ \citep{fontana_fraction_2009,wild_evolution_2016,forrest_zfourge_2018,clausen_3d-dash_2024}. These findings imply that high-mass galaxies tend to quench \blue{earlier} than their low-mass counterparts, a phenomenon widely referred to as \textit{downsizing}. This mass-dependent quenching trend is expected to extend to $z > 3$ \citep{xie_first_2024}. However, limited by the sensitivity of pre-JWST observatories, the quiescent fraction of high-mass galaxies at $z > 3$ was roughly estimated to be $\sim 20\%$, while the low-mass quiescent population remained virtually undetected \citep{weaver_cosmos2020_2023}.

The advent of JWST has revolutionized our view of the early universe. The number of spectroscopically confirmed QGs at $z \gtrsim 3$ has surged in recent years. A significant population of massive QGs has been identified in deep, pencil-beam surveys \citep[e.g.,][]{valentino_atlas_2023,carnall_massive_2023,carnall_jwst_2024,kakimoto_massive_2024,weibel_rubies_2024,de_graaff_efficient_2024,nanayakkara_formation_2025,Zhang_QG_2025,baker_abundance_2025,Baker_700_2025,Rubies_graaff_2025,Ito_deepdive_2025}, sparking intense debate regarding whether their abundance exceeds theoretical predictions. Notably, a subset appears to have formed the bulk of their stellar mass at extremely early epochs ($z \gtrsim 6$, or even $z \gtrsim 10$) and subsequently remained quiescent \citep{glazebrook_massive_2024,de_graaff_efficient_2024}. The existence of such systems implies progenitors with exceptionally high star formation rates (SFRs), standing in tension with the apparent scarcity of extreme starbursts at these high redshifts. Furthermore, a population of low-mass, so-called ``mini-quenched" galaxies\symbolfootnote[1]{In this work, we define ``quiescent galaxies" as those with a time-since-quenching $t_q > 100,\mathrm{Myr}$, while ``mini-quenched" refers to systems with $30,\mathrm{Myr} < t_q < 100,\mathrm{Myr}$. See Section \ref{subsec:selection criteria} for details.} has recently been discovered at $z \sim 4-9$ \citep{Baker_miniquench_2025,looser_recently_2024,strait_extremely_2023,Baker_trouble_2025}. These galaxies, having recently ceased star formation, typically exhibit weak Balmer breaks and strong rest-frame UV emission, presenting a challenge to the standard downsizing paradigm. 

 To resolve these tensions, accurate determinations of the number density and physical properties of QGs are crucial. While a growing number of studies have recently estimated the abundance of high-redshift QGs \citep[e.g.,][]{valentino_atlas_2023, carnall_massive_2023, Zhang_QG_2025, Baker_700_2025, merlin_downsizing_2025}, most of these works rely primarily on NIRCam-based photometry. However, the lack of mid-infrared coverage introduces potential systematic biases. Recent studies have confirmed that mid-infrared data, probing the rest-frame near-infrared, is essential to accurately measure stellar masses of massive galaxies at $z > 5$~\citep{papovich_ceers_2023,wang_true_2025}. \blue{Theoretically, by probing the rest-frame near-infrared (e.g., $J$-band) at $z > 3$, MIRI provides vital constraints on dust attenuation. This capability impacts QG selection via the $UVJ$ diagram and specific star formation rate (sSFR) thresholds, potentially breaking the degeneracy between dusty star-forming galaxies and true QGs, thereby refining the measured number densities.}

In this study, we address these limitations by exploiting the deep, wide-area JWST/MIRI imaging (F770W and F1800W) from the PRIMER program to robustly characterize the QG population at $0.5 < z < 8$. Our analysis demonstrates that mid-infrared coverage is instrumental in refining the selection of high-redshift QGs. Specifically, at $z \sim 3-5$, MIRI helps to identify and exclude dusty star-forming contaminants that can mimic massive QGs. Furthermore, at $z \gtrsim 5$, we find that MIRI is pivotal for recovering a subset of QG candidates that are otherwise missed by near-infrared selection alone.



The paper is organized as follows. Section \ref{sec:Data and fitting} describes the observational data and reduction procedures. Section \ref{sec: sample selection} outlines our selection technique, presenting the selected QG sample. In Section \ref{sec: result}, we evaluate the impact of MIRI on SED fitting and classification, followed by the presentation of the evolution of QG number density, fraction and stellar mass function up to $z=8$. We discuss the implications of our findings in Section \ref{sec: discussion and conclusion} and summarize our conclusions in Section \ref{sec:summary}. Throughout this work, we assume a standard $\Lambda$CDM cosmology with $H_0 = 70~\mathrm{km~s^{-1}~Mpc^{-1}}$, $\Omega_m = 0.3$, and $\Omega_\Lambda = 0.7$. All stellar masses are calculated assuming a Chabrier initial mass function (IMF).

\section{Data and SED fitting method}
\label{sec:Data and fitting}

\subsection{Data}
\label{sec:Data}

In this study, we utilize the Public Release Imaging for Extragalactic Research (PRIMER, GO 1837, PI: James Dunlop), one of the largest and deepest surveys that provides JWST/NIRCam and MIRI imaging. The survey spans ten passbands: NIRCam F090W, F115W, F150W, F200W, F277W, F356W, F444W and F410M, plus MIRI F770W and F1800W. The imaging reaches deep depths (e.g., $3\sigma$ detection limit at a 0.2" fixed aperture is $\sim 29.7$ mag in F277W and $\sim 26.5$ mag in F770W). We also incorporated available imaging observations from other programs within the PRIMER fields; the details are identical to those reported in \cite{wang_true_2025} (hereafter W25).  This compilation is among the most comprehensive datasets using MIRI to date, providing a strong balance between field of view and multi-band coverage. Most galaxies in our dataset have photometric measurements in more than 30 bands, and the effective MIRI-covered area is 323.0 $\mathrm{arcmin}^2$. To mitigate contamination from Little Red Dots (LRDs; see Section~\ref{subsec:LRD contam}), we restrict the analysis to sources simultaneously covered by F115W, F150W, F200W, F277W, F356W, F444W, and F770W, resulting in a working area of 227 $\mathrm{arcmin}^2$.

We process all NIRCam and MIRI data with a customized JWST Calibration Pipeline (v1.13.4), which produces higher-quality images than the default pipeline. Source detection is performed using \texttt{SExtractor} v2.25.0 \citep{bertin_sextractor_1996} on a combined detection image constructed from F277W, F356W, F410M, and F444W. Photometry is measured in six circular apertures (0.2", 0.3", 0.4", 0.5", 0.7" and 1.0") and an elliptical Kron aperture; we apply aperture corrections and adaptively choose the best aperture for each source. \blue{We derive the detection completeness in F444W by comparing the galaxy number density in our field with those measured in the significantly deeper JADES survey.  Our analysis indicates that at $z\sim8$, the sample detection completeness is approximately 70\% for galaxies down to $\log(M_\star/M_\odot) \sim 9.0$. However, mere detection does not guarantee classification: as shown in Section \ref{sec: result}, the final QG sample selection is further constrained by the availability of MIRI coverage. The implications of this MIRI-driven selection incompleteness are discussed in detail in Section \ref{subsec:impact of MIRI}.} Full details of the reduction, source detection, and the multi-wavelength photometric catalog will be presented in a forthcoming paper (Sun et al., in prep).  Within the mass range $\log M/M_{\star}>9$, the photometric sample includes 14,830 sources at $0.5<z<8$.

\subsection{SED fitting method}

Photometric redshifts are computed using \texttt{EAZY} \citep{brammer_eazy_2008} for all sources with $\rm SNR_{F444W}>7$ and MIRI coverage, adopting the \texttt{sfhz\_blue\_13} template set. The dense band coverage provides reliable, high-quality photo-z estimates. The photometric-spectroscopic redshift comparison is presented in W25, yielding a normalized median absolute deviation of $\sigma_{\rm NMAD} = 0.017$ (see \cite{brammer_eazy_2008} for definition).

We perform SED fitting using \texttt{BAGPIPES}. For sources with MIRI coverage but SNR below $2\sigma$, we adopt a $3\sigma$ flux upper limit. We adopt the stellar population synthesis model of \cite{bruzual_stellar_2003}, with $age \in \rm[0.03,10]~Gyr$ and metallicity $Z/Z_\odot \in [0,2.5]$. We use the dust attenuation model of \cite{calzetti_dust_2000}($A_V \in [0,5]$), nebular emission constructed by \cite{byler_nebular_2017}($\rm{log}U \in [-5,-2]$) and a delayed exponentially declining star formation history($\tau \in [0.01,10]~\rm{Gyr}$).
We do not include AGN template in fitting. When a spectroscopic redshift is available, the redshift is fixed to the spectroscopic value during fitting.

\section{SAMPLE of QUIESCENT GALAXIES}
\label{sec: sample selection}

\subsection{Selection methods}
\label{subsec:selection criteria}


\blue{The identification of QG relies primarily on two established methods: rest-frame UVJ color-color diagram and thresholds based on sSFR (typically calculated by SFR averaged over past 100 Myr). At low redshift, these techniques are largely interchangeable, yielding statistically consistent samples dominated by evolved, long-quiescence galaxies \citep[e.g.,][]{muzzin_evolution_2013,straatman_substantial_2014, van_der_wel_3d-hstcandels_2014}.}

\blue{However, the landscape changes significantly at high redshift. A rising population of young quiescent galaxies (i.e., post-starburst galaxies) are found to fall outside the traditional UVJ selection wedge \citep{stevenson_primer_2025,Zhang_QG_2025,baker_abundance_2025},  necessitating empirical extensions to the UVJ criteria to capture them. In contrast, the sSFR-based selection --- specfically the dynamic threshold of $\mathrm{sSFR}<0.2/t_{\mathrm{Hubble}}(z)$ --- offers a more physically motivated definition. By scaling with the age of the Universe, this criterion naturally adapts to the evolving star-formation main sequence. Consequently, we adopt an sSFR threshold as our primary definition for high-redshift QGs. This approach effectively captures the diverse quiescent population at early epochs and aligns with standard practices in recent photometric studies \citep[e.g.,][]{carnall_massive_2023,Baker_700_2025}.}

\blue{We implement a three stage procedure to construct a robust sample of QGs:}
\blue{First, we apply a relaxed UVJ color to filter out majority of dust-rich galaxies and star-forming galaxies while retaining potential QG candidates. The selection criteria are defined as:}

\begin{align}
(V - J) &< 1.8 \\
(U - V) &> 0.88 \cdot (V - J) + 0.29 
\end{align}

\blue{These loose boundaries are specifically designed to be inclusive; they encompass all spectroscopically confirmed QGs reported in recent high-redshift studies \citep{baker_abundance_2025,Zhang_QG_2025}, ensuring that young or post-starburst systems are not prematurely discarded, while effectively rejecting severe dusty contaminants.}

\blue{From the pre-selected candidates, we identify true QGs using our primary criterion based on the sSFR. We adopt the widely accepted redshift-dependent threshold:}
\begin{align}
    \mathrm{sSFR}_{100}<0.2/t_{\mathrm{Hubble}}(z)
\end{align}

\blue{Furthermore, to explore the population of recently quenched galaxies (``mini-quenched" galaxies), we also identify sources satisfying $\mathrm{sSFR}_{30\mathrm{Myr}} < 0.2/t_{\mathrm{Hubble}}(z)$ (averaged over the last 30 Myr). This supplementary selection allows us to evaluate the full extent of the quenching phenomenon, including its most rapid phases.}

\subsection{Reduce Contamination}
\label{subsec:LRD contam}

\blue{ While sSFR serves as our primary standard for identifying QGs, samples derived purely from photometric SED fitting are susceptible to systematic contamination. A critical concern in the JWST era is the prevalent population of ``Little Red Dots" (LRDs) at $z \gtrsim 3$ \citep{labbe_uncover_2023,li_little_2024,greene_uncover_2024,matthee_little_2024}. These compact, broad-line objects exhibit V-shaped SEDs that can closely mimic the photometric signatures of QGs. Crucially, their ubiquitous broad emission lines (e.g., [O III], H$\alpha$) can simulate a strong Balmer break, leading to catastrophic photometric redshift errors \citep{desprez_cdm_2024, barro_extremely_2024} and causing a QG-like sSFR fitting result.}


%
\blue{To mitigate this specific contamination, we implement a hybrid filtering strategy adapted from \cite{kocevski_rise_2024}, \cite{labbe_population_2023} and \cite{Rinaldi_dot_2025}, which prioritizes morphological constraints, supplemented by spectral shape analysis:}

\begin{align}
    & \text{compactness} = \mathrm{Ap}_{0.4}/\mathrm{Ap}_{0.2} < 1.7 \\
    & \beta_{\mathrm{opt}} > 0 \\
    & -2.8 < \beta_{\mathrm{UV}} < -0.37 \\
    & F150W - F200W < 0.8~\mathrm{mag} \\
    & F277W - F444W > 0.7~\mathrm{mag}
\end{align}

\noindent \blue{ where $\beta_{\mathrm{UV}}$ and $\beta_{\mathrm{opt}}$ are the rest-frame UV and optical spectral slopes, respectively. To calculate the slopes, we adopt the same redshift-dependent filter selection as used in \cite{kocevski_rise_2024}. The best-fit parameters are derived using the \textit{scipy.optimize.curve\_fit} algorithm, with the uncertainties calculated as the square root of the diagonal elements of the resulting covariance matrix. }

\blue{\textbf{Step 1: Morphological Screening.} We first evaluate the morphology of all candidates. Since LRDs are strictly defined as compact, point-like sources, any galaxy that exhibits extended structure (i.e., fails the LRD compactness criteria Eqa.4) is immediately retained as a robust QG candidate, regardless of its SED shape.}

\blue{\textbf{Step 2: Spectral V-Shape Check} (For Compact Sources Only). For sources identified as morphologically compact, we further scrutinize their SEDs for the characteristic V-shaped profile {(Eqa.5-8). The method used by Eqas 5 and 6 relies on photometric redshift derived by \texttt{EAZY}. However, the photometric redshift estimate for LRDs suffer from large uncertainties due to their abnormal SEDs. Therefore, we also use Eqas 7 and 8 to remove possible LRD contamination.} This assessment depends on the photometric quality:}

 \blue{\textit{High-SNR Case:} If the photometry is sufficient to confirm a V-shaped profile at the $>1\sigma$ level, the compact source is classified as an LRD and removed. Conversely, if the data rule out a V-shape, the compact source is retained.}
 
 \blue{\textit{Low-SNR Case:} If the data lack sufficient depth to unambiguously determine the spectral shape, we take a conservative approach: all compact sources with ambiguous SEDs are removed, as QGs at high redshift typically have extended structure. }


\blue{ 
The step of LRD filtration step is only applied to candidates at $z > 3$, as the LRD population is negligible at lower redshifts.
}

Brown dwarfs can also act as contamination in QG candidates at $z>5$. To mitigate this, we exclude all sources satisfy $F150W-F200W<0.25$, consistent with the brown dwarf selection criteria from \cite{langeroodi_browndwarfs_2023}.

\blue{Beyond specific contaminant populations, we also enforce a general quality control cut. We exclude any QG candidates where the best-fit model deviates from the observed photometry by more than $2\sigma$ in the key rest-frame optical/near-infrared bands (F356W, F444W, or F770W). Such significant residuals suggest that the source cannot be adequately described by standard stellar and nebular templates, potentially indicating incorrect redshifts or unmodeled AGN components.}

\subsection{galaxies in UVJ diagram}

\begin{figure*}[htbp]
    \centering \includegraphics[width=0.9\textwidth]{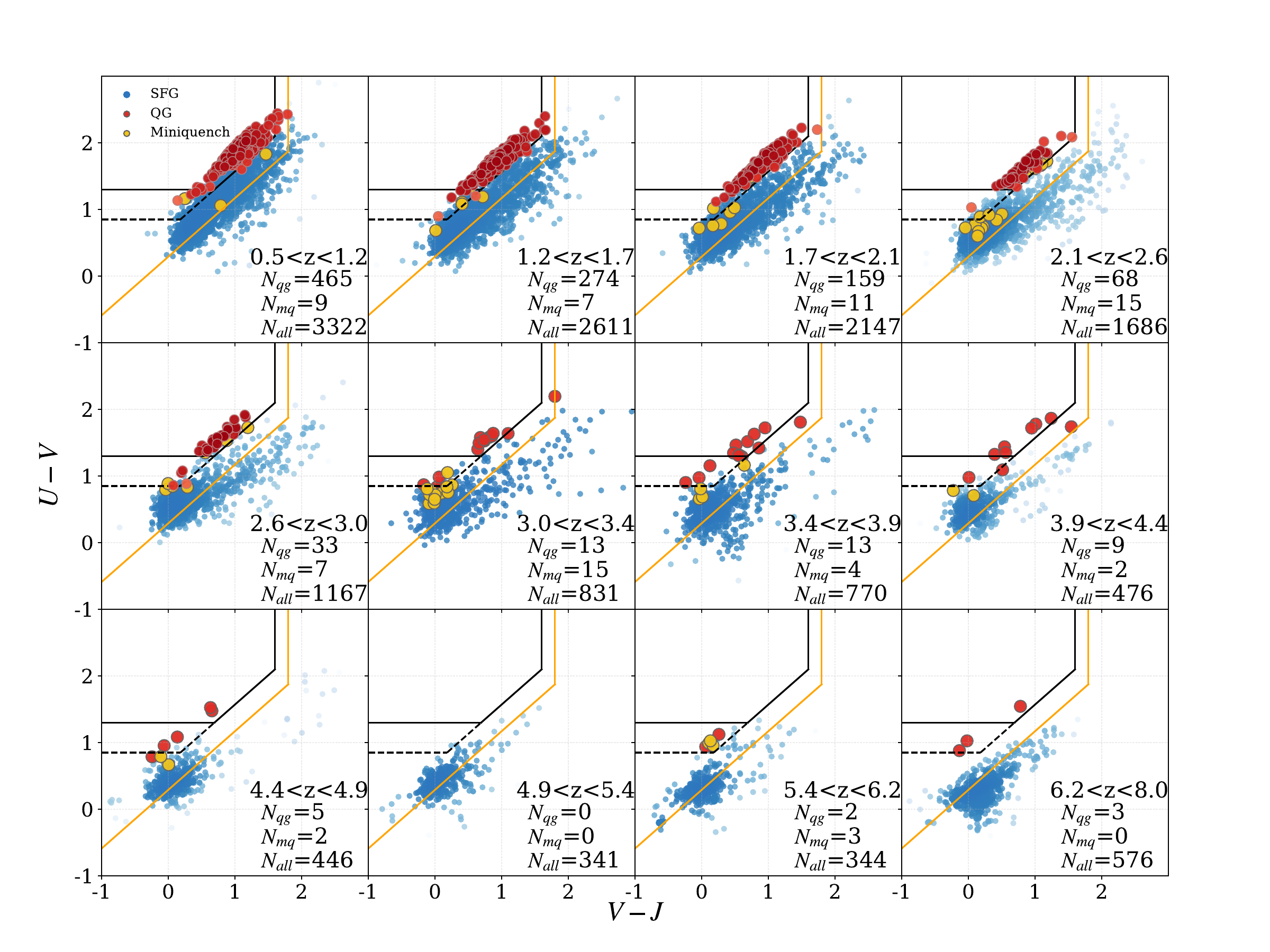}
    \caption{\textbf{The UVJ color-color diagram for the photometric sample with $M_{\star}>10^{9.0}M_{\odot}$.} The comoving volume is nearly constant across all redshift slices (except for the highest-redshift bin), allowing for a direct comparison of the number density evolution of QGs and SFGs. The orange line denotes our inclusive pre-selection boundary (defined in Eqs. 1 and 2). The solid black line indicates the standard UVJ criterion proposed by \cite{williams_detection_2009}, while the dashed black line shows an extended criterion (lowering the $U-V$ threshold to 0.85). This extended cut is designed to include high-redshift PSBs, though it introduces substantial contamination from SFGs. \review2{This diagram does not include contamination removed in Sec \ref{subsec:LRD contam}}. }
    \label{fig:UVJ selectin}
\end{figure*}

\blue{We identify a total of {45} QG candidates at $3<z<8$. The mass distribution of this sample comprises 4 sources with $\log M_{\star}/M_{\odot}<9.5$, {30} sources with $9.5<\log M_{\star}/M_{\odot}<10.6$, and 11 massive galaxies with $\log M_{\star}/M_{\odot}>10.6$. Additionally, we identify \review2{26} mini-quench galaxies, with masses predominantly cluster in the range $\log M_{\star}/M_{\odot} \sim 9.2-10.2$. }

\blue{Figure~\ref{fig:UVJ selectin} presents the distribution of these quiescent and mini-quench galaxies on the UVJ plane. To facilitate a visual comparison of number density evolution, the redshift bins in this figure (with the exception of the highest-redshift bin) are defined to enclose equal comoving columns.}

\blue{At low-redshifts, the distnction between QGs and SFGs is well-approximated by the classic boundary: $(U-V)>1.3; (U-V)>0.88\times (V-J)+0.69; (V-J)<1.6$ (indicated by the black dashed line). However, as introduced in section \ref{subsec:selection criteria}, strictly adhering this boundary at high redshift would result in significant incompleteness. The majority of QGs at $z\gtrsim4$ locate between $0.85<U-V<1.3$, however, lowering the color cut to $U-V=0.85$ would inevitably introduce substantial contamination from SFGs, further motivating our adoption of the sSFR-based classification.}


\blue{On the UVJ plane, the mini-quench population primarily occupies the region immediately below the young QGs, overlapping considerably with the star-forming sequence. This placement is consistent with their recent suppressed star-formation, which results in bluer UV colors compared to fully evolved QGs. Quantitatively, their relative abundance evolves strongly with redshift. At $z<3$, mini-quench galaxies are a minor population, numbering less than 20\% of the QG sample. However, at $z>3$, the number density becomes comparable to that of the QG population. To maintain consistency with literature definitions, we do not include these mini-quench candidates into our QG sample. However, including these transitional objects would enlarge the sample size at $z>3$ by a factor of $\sim 2$ ($\approx0.3~\mathrm{dex}$ on the deived number density).}

\section{Result}
\label{sec: result}
\subsection{The importance of MIRI in QG classification}
\label{subsec:MIRI influence}


Previous studies indicate that the absence of JWST/MIRI data can lead to overestimated stellar masses for galaxies at redshifts $z > 5$ (W25; \citealt{papovich_ceers_2023,leung_exploring_2024}), which may artificially inflate the number density of high-mass QGs at these epochs. In addition, missing MIRI photometry, which probes the rest-frame J band at $z>3$, can bias $V-J$ color and sSFR measurements and thus affect QG classification.



To quantify the impact of MIRI data on derived physical properties and integrated QG number densities, we performed a comparative analysis by running two parallel sets of SED fitting with \texttt{BAGPIPES}: one utilizing the full photometric dataset (the \texttt{MIRI} run) and a control run excluding MIRI photometry (the \texttt{no-MIRI} run). In both cases, redshifts were fixed to the values derived from the MIRI-inclusive catalog, as MIRI photometry has been shown to add minimal constraints to $z_{\mathrm{phot}}$ when high-quality Hubble Space Telescope (HST) and JWST/NIRCam data are present \citep{stevenson_primer_2025}. \blue{In Section \ref{sec: sample selection}, we constructed our parent sample down to $\log (M_\star/M_\odot) = 9.0$. However, for our analysis of integrated number densities, we adopt a conservative threshold of $\log (M_\star/M_\odot) > 9.5$ across all redshift bins. This choice is motivated by two key factors: first, it ensures high sample completeness ($>95\%$) up to $z\sim8$, thereby mitigating uncertainties from corrections; second, it facilitates a direct and consistent comparison with the majority of previous studies \citep[e.g.,][]{Baker_700_2025,russell_QG_2025,valentino_atlas_2023}, which typically adopt a similar mass limit.}

\begin{figure*}
    \centering
    \includegraphics[width=0.8\textwidth]{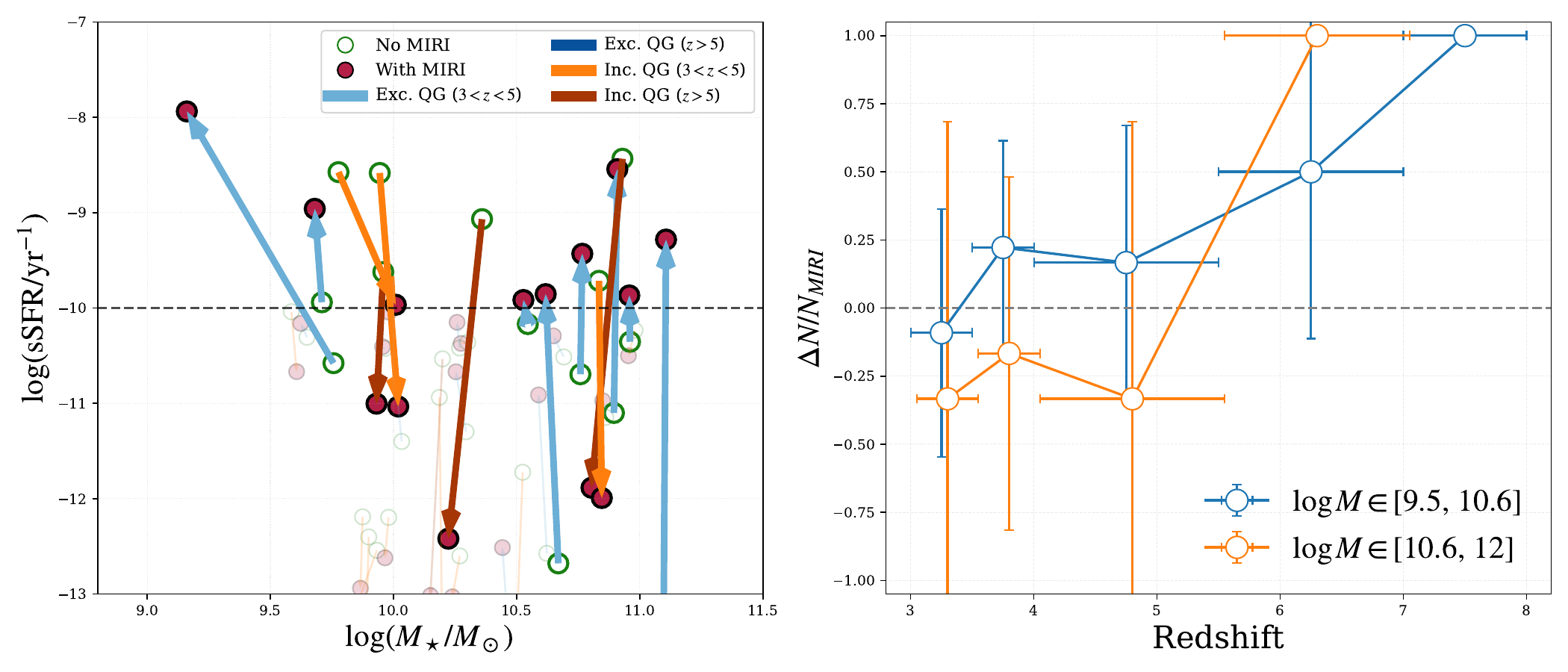}
    \includegraphics[width=0.8\textwidth]{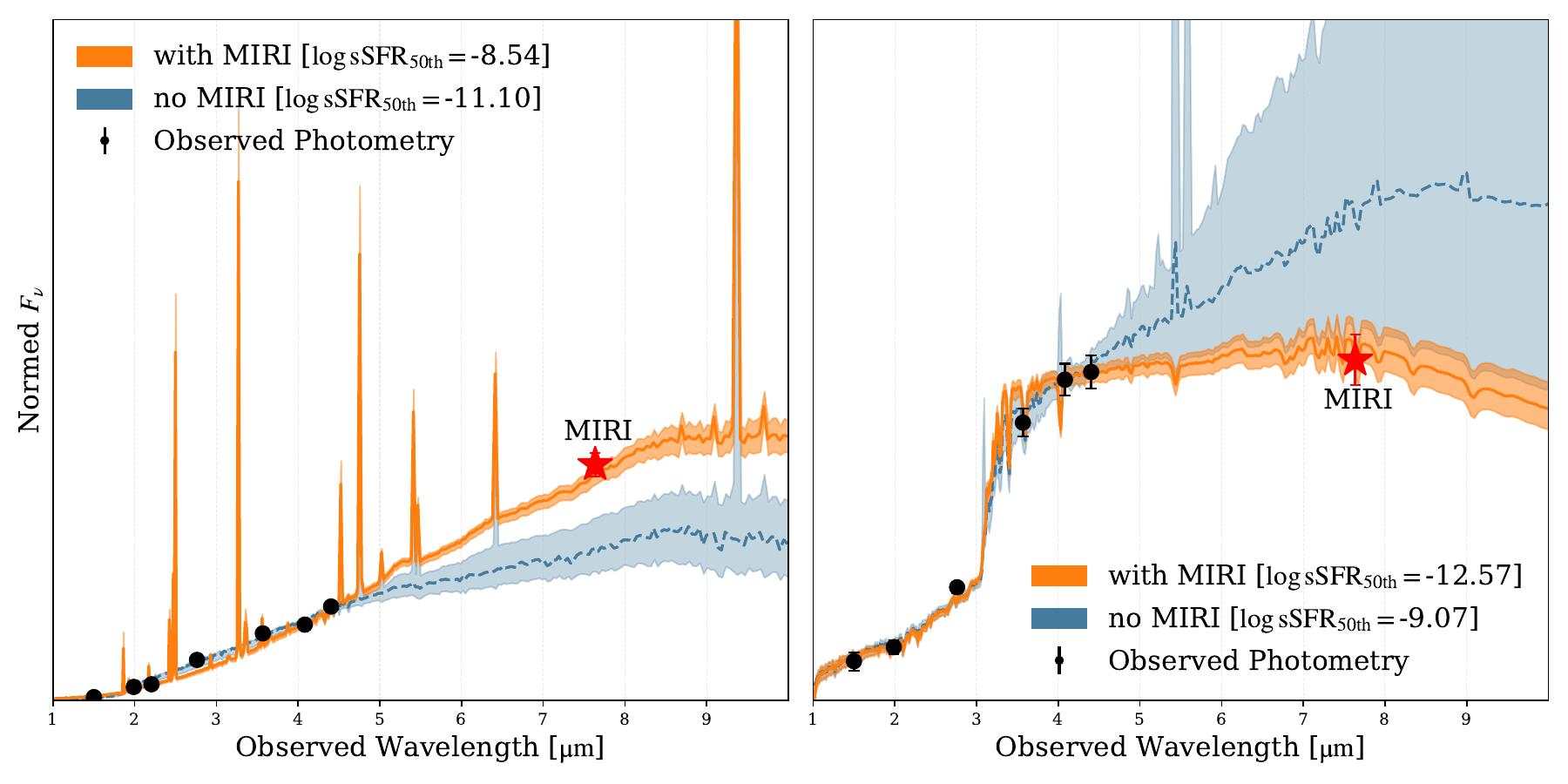}
    
    \caption{\blue{\textbf{Impact of MIRI photometry on quiescent galaxy selection.} \textbf{Top-left:} Migration of galaxies in the sSFR–$M_{\star}$ plane upon the inclusion of MIRI photometry. Orange arrows highlight sources reclassified from Star-Forming (SFG) to Quiescent (QG), while blue arrows indicate the reverse (QG $\rightarrow$ SFG). Sources with unchanged classifications are shown with reduced opacity. \textbf{Top-right:} The relative change in QG number density as a function of redshift across different stellar mass bins. Error bars represent Poisson uncertainties. \textbf{Bottom panels:} Two illustrative examples where MIRI photometry is decisive for correct classification. \textbf{Lower-left:} A dusty star-forming galaxy where the addition of MIRI data resolves SED degeneracies, leading to a correct fit. \textbf{Lower-right:} A quiescent galaxy at $z=7.2$ (spectroscopically confirmed by \citealt{weibel_rubies_2024}) that was misidentified as star-forming without MIRI constraints, demonstrating the critical role of MIRI coverage in identifying high-redshift QGs.} }
    \label{fig:ssfr comparision}
\end{figure*}

\blue{Based on the sSFR derived from the two fits, the MIRI run identifies 41 QGs with $\log (M_\star/M_\odot)>9.5$. Subdividing the sample at $\log (M_\star/M_\odot) = 10.6$ reveals 11 high-mass and 30 low-mass galaxies. In contrast, the no-MIRI run also yields 41 QGs but with a shift towards higher masses (15 high-mass, 26 low-mass), showing partial overlap with the MIRI-selected sample. Figure \ref{fig:ssfr comparision} illustrates how the inclusion of MIRI data shifts galaxies on the sSFR-mass plane and tracks the evolution of the fractional difference ($(N_{\mathrm{MIRI}}-N_{\mathrm{noMIRI}})/N_{\mathrm{MIRI}}$) in QG number density with redshift. Two representative examples demonstrating how MIRI constrain alters the best-fit parameters are also presented.  }

\blue{We find that the impact of MIRI on QG selection depends strongly on stellar-mass and redshift. At the high-mass end, MIRI primarily acts to purge contaminants. Of the 15 high-mass candidates in the no-MIRI sample, 5 are rejected after incorporating MIRI data. All of them locate at $z<5$. These contaminants are nearly equally divided into two categories: 3 are dust-rich galaxies revealed to have high sSFRs ($\log \mathrm{sSFR}\gtrsim-9.5$; see Fig.~\ref{fig:ssfr comparision}, lower left), and 2 are marginal sources that shift to the regime near the classification boundary (within 0.1 dex) upon the inclusion of MIRI constraints.}

\blue{In contrast, at the low-mass end, MIRI becomes crucial for recovering genuine QGs that are otherwise missed by NIRCam-only photometry. This effect is dominant at $z>5$: among the 3 low-mass QGs identified with MIRI at $z>5$, 2 required MIRI data for confident selection. For the single identified QG with $\log M_\star/M_\odot>10.6$ at $z>5$, MIRI is also needed to determine its properties. This recovery addresses the severe degeneracy in NIRCam-only fits, where parameter solutions are often multimodal; quiescent solutions typically occupy only the tail of the posterior probability distribution and are marginalized out (see Fig.~\ref{fig:ssfr comparision}, lower right). MIRI constraints effectively rule out these star-forming solutions, thereby enhancing the completeness of the QG sample.}

{While MIRI currently offers a critical lever for breaking these degeneracies, additional NIRCam observations—particularly those utilizing medium-band filters—could represent a potential alternative strategy. In principle, such data would help constrain strong emission lines and provide a finer sampling of the Balmer break profile, potentially aiding in the identification of contaminants or missing QGs in fields lacking MIRI coverage. However, the extent to which more NIRCam photometry can effectively substitute for the constraints provided by MIRI requires further investigation.}

\subsection{Number density and Fraction of Quiescent galaxies}

\begin{figure*}
      \centering
    \includegraphics[width=0.8\textwidth]{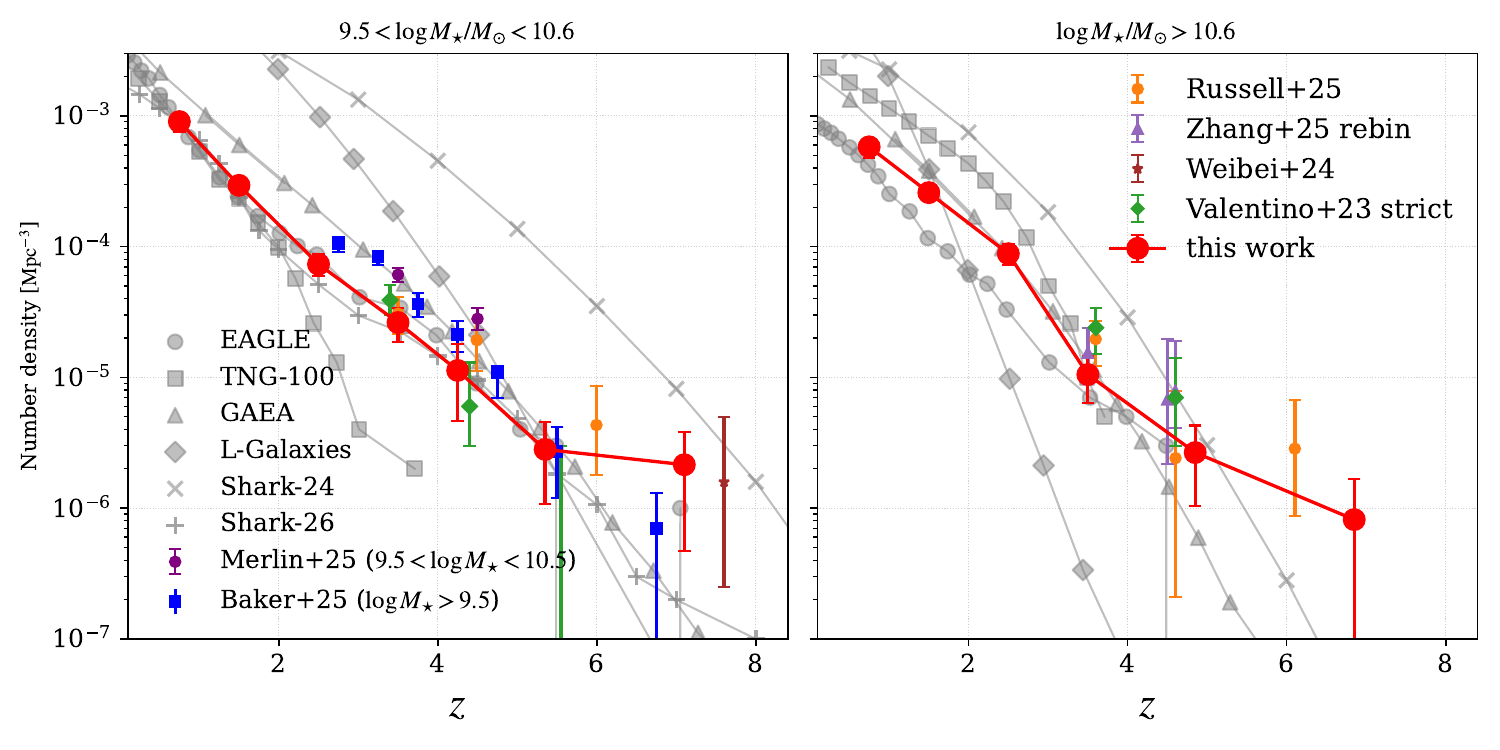}
    \caption{ \textbf{The mass-dependent number density evolution of QGs.} Error bars represent combined effects of Poisson noise and cosmic variance; for low-redshift data points, relative errors are approximately 10\% and thus not visible. Observational results from \cite{valentino_atlas_2023,russell_QG_2025,Zhang_QG_2025,Baker_700_2025,merlin_downsizing_2025} are included for comparison. To ensure consistency, data from \cite{Zhang_QG_2025} have been recalculated for the mass range $\log M_{\star}/M_{\odot} > 10.6$. Unless otherwise noted, all comparison works report number densities within the same mass range indicated in the panel titles. Small redshift offsets ($\Delta z = 0.1$) are applied to some data points to minimize overlap. We compare our results with predictions from EAGLE, Illustris-TNG, GAEA, L-Galaxies, and Shark, adopting a uniform selection threshold of $\mathrm{sSFR}<0.2/t_{\rm Hubble}(z)$ for the models.}
    \label{fig:density evolution}
\end{figure*}

We present the evolution of integrated number density of robust QGs in Figure~\ref{fig:density evolution}. We divide the sample at $10^{10.6} M_\odot$ into low-mass and high-mass subsamples. {The cosmic variance in the statistics is estimated following \cite{moster_cosmic_2011}. The total error is obtained by adding the cosmic and Poisson errors in quadrature.}
 Both subsamples show a rapid decline in number density with increasing redshift at $z<4$, but their evolutionary trajectories diverges at $z>4$. 

The number density of high-mass QGs decreases rapidly with increasing redshift, reaching \blue{$n \sim 10^{-5}~\mathrm{Mpc^{-3}}$ at $3 < z < 4$ and falling to $n < 10^{-6}~\mathrm{Mpc^{-3}}$} at $z>6$. By contrast, the number density of low-mass QGs remains nearly constant from $z=4$ to $z=8$, at approximately {$n \sim2\times10^{-6}~\mathrm{Mpc^{-3}}$}.



\begin{figure*}[htbp]
    \centering
    \includegraphics[width=0.9\textwidth]{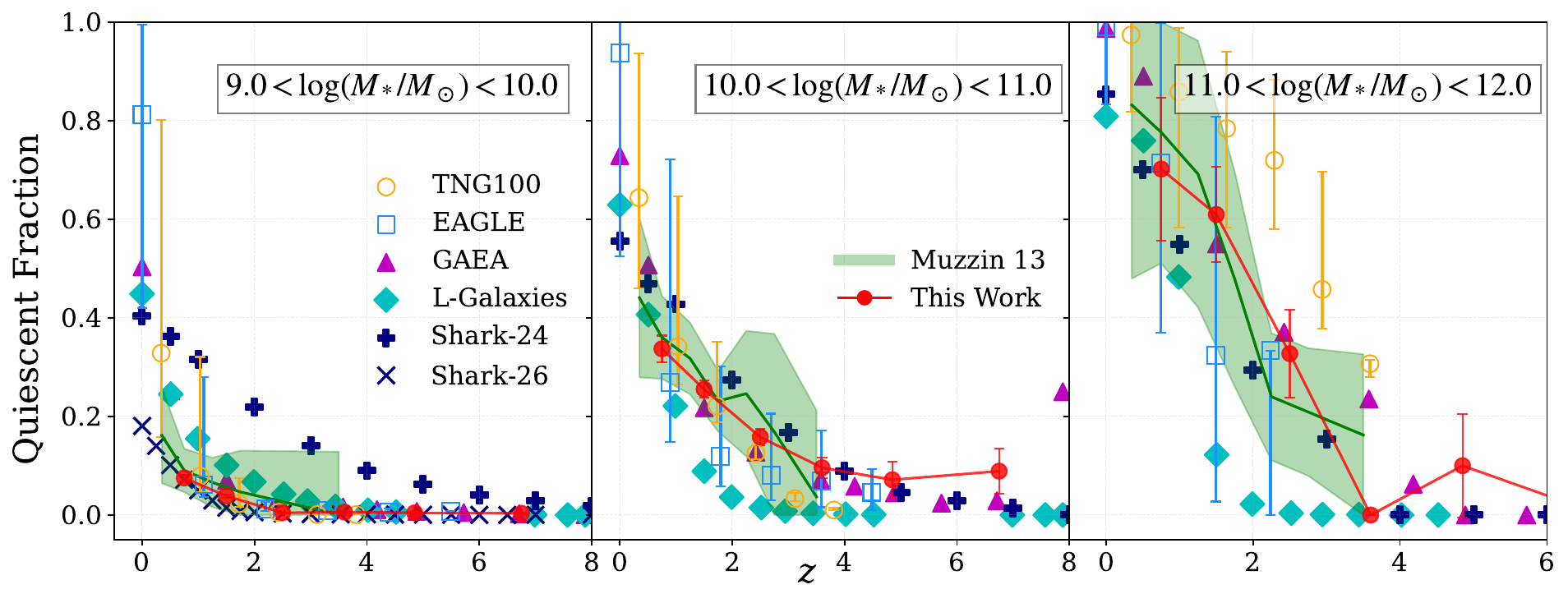}

    \caption{\textbf{The redshift evolution of the QG fraction in different stellar mass bins.} Predictions from EAGLE, Illustris‑TNG, GAEA, L‑Galaxies, and Shark, along with observational data from \cite{muzzin_evolution_2013}, are plotted for comparison. We include the $9<\log M_{\star}/M_{\odot}<9.5$ bin to match the mass range of \cite{muzzin_evolution_2013}, noting that completeness issues have a reduced impact on the fraction compared to number density. Model QGs are selected using $\mathrm{sSFR}<0.2/t_{\mathrm{Hubble}}(z)$. While discrepancies exist, the best-performing theoretical models reasonably reproduce the QG fraction evolution up to $z=8$.}
    \label{fig:mass bin fraction}
\end{figure*}

\blue{We further investigate the evolution of the quenched fraction as a function of stellar mass and redshift (Fig.~\ref{fig:mass bin fraction}). We include the $9.0 < \log(M_{\star}/M_{\odot}) < 9.5$ bin to align with the mass range of \cite{muzzin_evolution_2013}, noting that incompleteness effects largely cancel out in the fraction ratio compared to absolute number densities. At high redshift ($z>4$), the QG fraction remains consistently low ($<2\%$) across the full mass range, with only a marginal increase to $\sim7\%$ for galaxies with $10 < \log(M_\star/M_\odot) < 11$. However, toward lower redshifts, the evolution diverges strongly depending on mass. The high-mass quenched fraction surges from 5\% at $z=4$ to 40\% by $z=2$. In contrast, the low-mass fraction grows sluggishly, remaining below 20\% even in the local Universe.}

\subsection{Stellar Mass Function of Quiescent Galaxies}

\begin{figure*}
    \centering
    \includegraphics[width=0.8\linewidth]{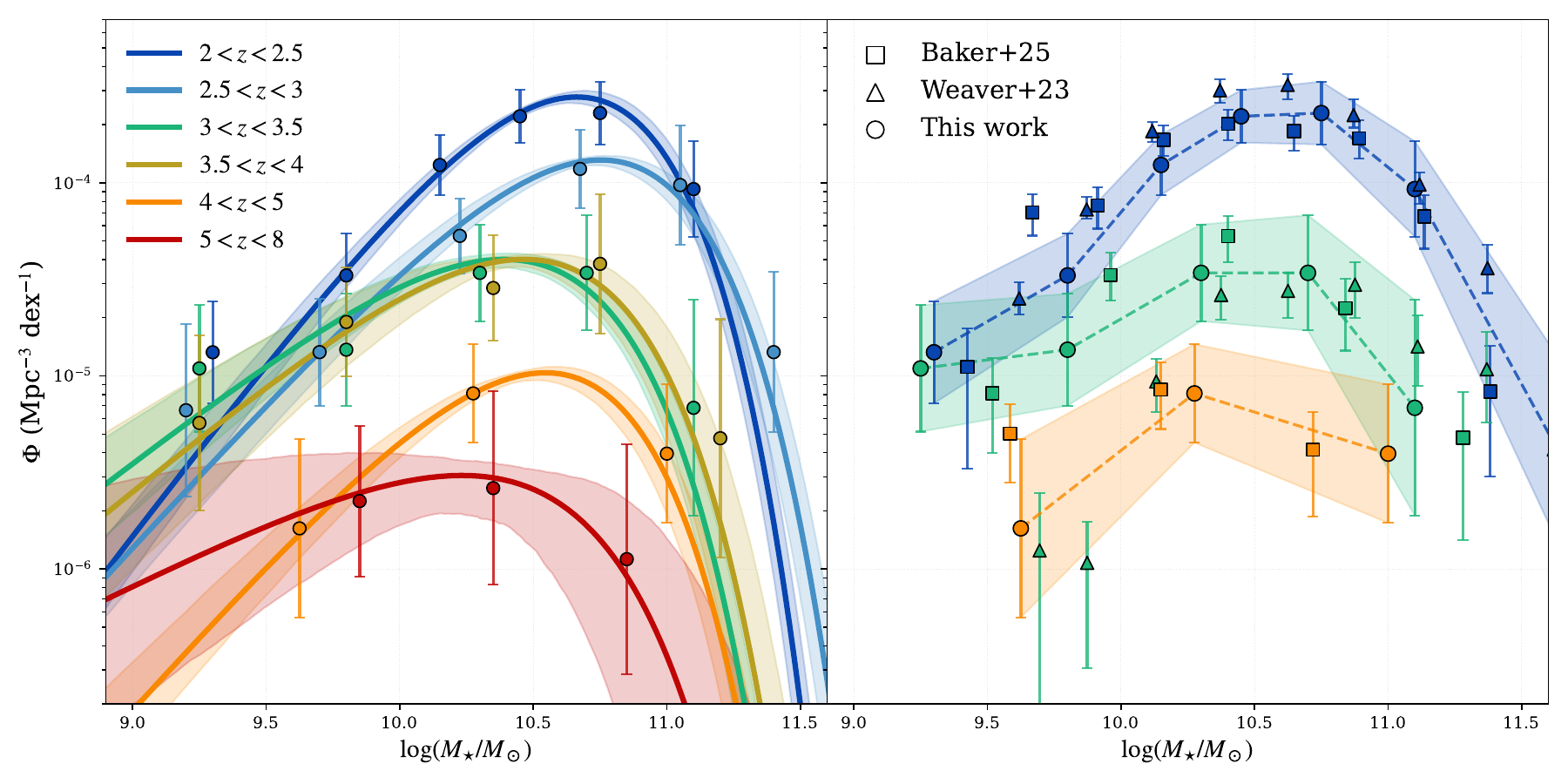}
    \caption{ Left panel: \textbf{The Stellar Mass Function (SMF) of QGs at $2<z<8$, fitted with a single Schechter function.} Galaxies with $\log (M_\star/M_\odot)=9-9.5$ are included in the fit only in redshift bins where the sample is complete. Data points and the best-fit Schechter functions are shown. The fitting methodology follows the procedure described in \cite{sun_protocluster_2024}. Right panel: Comparison of our observations with \cite{Baker_700_2025} and \cite{weaver_cosmos2020_2023} in three redshift ranges: $z\sim 2-2.5$, $z\sim 3-3.5$, and $z\sim 4-5$.}
    \label{fig:highz_SMF}
\end{figure*}

\blue{Figure \ref{fig:highz_SMF} shows the stellar mass function (SMF) of QGs at $z>2$. We set redshift bins of \review2{2–2.5, 2.5–3, 3–3.5,3.5–4, 4–5, and 5–8.} Galaxies with $\log (M_\star/M_\odot)=9-9.5$ are included in the SMF fitting only when they are complete at the corresponding redshift bins.}

\blue{We fit the SMF with the single
Schechter function:}

\begin{align}
    \Phi d(\log M) = \ln(10)\times \exp(-10^{\log M-\log M^*}) \nonumber \\  \times [\Phi_1^*(10^{\log M-\log M^*})^{\alpha_1+1}]d (\log M)
\end{align}

\blue{While the double Schechter function is preferred for fitting the SMF at low redshift for capturing the distinct high- and low-mass components, it is ill-suited for our high-redshift sample due to the limited statistics at the low-mass end. Consequently, we adopt a single Schechter function, which provides a robust description of our sample and facilitates direct comparison with similar high-redshift studies \citep{Baker_700_2025,weaver_cosmos2020_2023}.  }

The evolution of the SMF reveals a contrast between mass ranges. We observe a rapid decline in the number density of high-mass QGs with increasing redshift, mirroring the strong evolution of their integrated number density. {The only exception is that the SMF at 3.5$-$4 does not show a significant decline compared to 3$-$3.5, which is due to an overdensity of QGs in a $z = 3.98$ protocluster~\citep{tanaka_protocluster_2024, bigfoot_sun_2025}. Conversely, the low-mass end of the SMF exhibits remarkable stability. Specifically, provided the definition of ``low-mass" is restricted to $\log (M_\star/M_\odot)\lesssim10.3$, the SMF shape at low masses remains nearly identical between \review2{$4<z<5$} and \review2{$5<z<8$}, aligning with the observed plateau in low-mass QG number density at $4<z<8$. A similar stagnation in the number density of low-mass QGs growth is also evident at lower redshifts $2<z<4$~(which is also reported by \citealt{Baker_700_2025}). While this lower-redshift plateau is $\sim 0.5 $ dex higher than the one at $z>4$, our findings suggest that the buildup of QG in the low-mass regime stalls significantly over long cosmic epochs, showing little effective growth within $2<z<8$.}

\section{Discussion}
\label{sec: discussion and conclusion}

In this work, we estimate the number density of QGs at $0.5<z<8$ using multi-wavelength data, including MIRI observations. We find a mass-dependent evolution of the number density, providing new constraints on galaxy quenching at high redshift.

\subsection{Comparison with previous work}

\blue{In the JWST era, numerous studies have estimated the number density of QGs using both photometric samples (e.g., \citealt{valentino_atlas_2023}, with partial MIRI coverage in \citealt{Baker_700_2025,russell_QG_2025,stevenson_primer_2025}) and spectroscopic confirmations, particularly for high-mass galaxies \citep{weibel_rubies_2024,Zhang_QG_2025}.}

 \blue{At high-mass end, our results show great agreement with the spectroscopic constraints from \cite{Zhang_QG_2025} at $z\sim3$. When compared to photometric studies, our high-mass number densities are broadly consistent with \cite{russell_QG_2025} and \cite{valentino_atlas_2023} within the uncertainties, yet we observe a systematic offset, with our values being lower by $\sim 0.1-0.2~\mathrm{dex}$. This systematic difference reinforces our earlier conclusion that photometric selection without MIRI constraints is prone to contamination from dusty star-forming galaxies, leading to slightly inflated number densities. At higher redshifts ($z>4$), while estimates across different works converge, the constraints remain limited by small sample statistics and large uncertainties.}

At the low-mass end and $z \sim 3-6$, our identified QG abundance is generally consistent with previous literature. Extending to $z > 6$, our results remain in good agreement with estimate based on spectroscopic samples~\citep{weibel_rubies_2024}. However, we find \review2{relatively} higher number densities at $z > 6$ compared to \cite{Baker_700_2025}, \review2{although the difference remains within the $1\sigma$ uncertainties.} Notably, this excess exists even though our integration range ($9.5 < \log M_\star/M_\odot < 10.6$) is narrower than the lower limit threshold used in their analysis ($>10^{9.5} M_\odot$). This difference supports our argument that deep MIRI photometry is crucial for recovering low-mass, high-redshift QGs that might otherwise be missed. 




\blue{Conversely, \cite{merlin_downsizing_2025} reported significantly higher number densities for low-mass QGs than found in this work. We attribute this discrepancy primarily to differences in the definition of ``quiescence." Their sample likely includes a substantial fraction of galaxies with short quenching timescales ($\tau_{\mathrm{quench}} \lesssim 100~\mathrm{Myr}$), which we categorize separately as ``mini-quench" galaxies. As shown in Figure~\ref{fig:UVJ selectin}, the abundance of these transitional galaxies at high redshift is comparable to that of the rigorous QG population. Consequently, the inclusion or exclusion of this population can easily account for a $\sim 0.2-0.3~\mathrm{dex}$ offset in derived number densities.}

{We compare our SMFs with those from \cite{Baker_700_2025} and \cite{weaver_cosmos2020_2023} across three redshift bins: \review2{$z=2-2.5$, $3-3.5$, and $4-5$}. Overall, our results show broad agreement with previous studies, though notable discrepancies exist at both mass ends.}

{At the low-mass end and $z > 3$, our number densities are substantially higher than those reported by \cite{weaver_cosmos2020_2023}. This is likely attributable to differences in quiescent galaxy selection criteria, which is also noted by \cite{Baker_700_2025}.}

{At the high-mass end, we observe a lack of massive QGs compared to literature values, especially at $\log M_\star/M_\odot>11$. This discrepancy likely arises from a combination of factors: cosmic variance, the effect of MIRI decontamination, and the overestimation of stellar masses due to lacking MIRI coverage \citep{papovich_ceers_2023,wang_true_2025}. While the statistical significance of MIRI-driven corrections is currently obscured by the cosmic variance inherent to small fields, our findings suggest that current measurements of the high-mass end of the SMF of QGs might be systematically overestimated.}

{We note that current analysis is still limited to small samples, future wide-area surveys incorporating MIRI photometry (and/or suitable NIRCam medium-band coverage) will be essential 
to put stringent constraints on the SMF of QGs.} 

\subsection{Comparison with galaxy formation models}
We compare our results with six galaxy formation models, including two cosmological hydrodynamical suites:
\begin{itemize}
    \item the IllustrisTNG simulation  \citep[TNG;][]{pillepich_first_2018, marinacci_first_2018, naiman_first_2018, nelson_first_2018, springel_first_2018}. We use the TNG100-1, with box size of $L_{box}=100.7 \mathrm{Mpc}$.
    \item the Virgo Consortium's Evolution and Assembly of Galaxies and their Environments \citep[EAGLE;][]{schaye_eagle_2015}. We use the Fiducial\_models..RefL0100N1504, with box size of $L_{box}=100 \mathrm{Mpc}$.
\end{itemize}
and four semi-analytical models:
\begin{itemize}
    \item the GAlaxy Evolution and Assembly model \citep[GAEA;][]{de_lucia_tracing_2024, hirschmann_galaxy_2016, xie_uence_2020}, with box size $L_{box}=500~\mathrm{Mpc}$.
    \item L-Galaxies \citep{Henriques_formation_2015,pei_simulating_2024}, with box size $L_{box}=676~\mathrm{Mpc}$.
    \item Shark2024 \citep{lagos_shark_2018, lagos_quenching_2024}, with box size $L_{box}=376~\mathrm{Mpc}$. 
    \item Shark2026 \citep{Oxland_quenching_2026}. The Shark2026 is an updated version of the Shark2024. The only difference with respect to the version presented in \cite{lagos_quenching_2024} is the ram pressure stripping (RPS) proceeds in a slow fashion, instead of the RPS invoked by \cite{lagos_quenching_2024}. Oxland et al. found that a slow RPS is required to reproduce the fractions of quenched galaxies in groups and clusters and as a function of cluster centric distance presented in \cite{Oxland_quenching_2024}. The difference between Shark2024 and 2026 in high-mass end is negligible, so we only present Shark(2026) data in low-mass end for clarity.
\end{itemize}

All referenced models adopt a Chabrier IMF. The hydrodynamical simulations, limited by their volume, have a single-detection limit of $\sim 10^{-6}~\mathrm{Mpc^{-3}}$, which is roughly consistent with our observational limits. In contrast, the SAMs utilize significantly larger simulation boxes, yielding much lower detection limits. While some models have demonstrated reasonable agreement with JWST number density measurements up to $z \sim 5$ \citep{lagos_quenching_2024,de_lucia_tracing_2024}, we extend this comparison to $z=8$. To ensure consistency, we apply a uniform selection criterion across all models, identifying quiescent galaxies using an sSFR threshold of $\mathrm{sSFR} < 0.2/t_{\mathrm{Hubble}}(z)$.


Regarding the fraction of QGs, most theoretical models predict values consistent with observations across all redshift bins, within uncertainties. However, a subset of models underestimates the QG fraction at high redshift; for instance, predictions from EAGLE and TNG extend only to $z \sim 4$ \review2{for galaxies with $\log M_\star/M_\odot>10$}, indicating a complete absence of QGs at earlier epochs in these simulations \citep[see also][]{weibel_rubies_2024,long_efficient_2023}. Notably, the \texttt{Shark} (2024) model appears to overestimate the low-mass QG fraction. This bias vanishes in the updated \texttt{Shark} (2026) prescription, suggesting that the precise treatment of environmental quenching is critical for reproducing the low-mass population. We caution, however, that the QG fraction depends on the abundance of both quiescent and star-forming galaxies, so discrepancies may not solely reflect errors in quenching prescriptions. Furthermore, as the intrinsic QG fraction approaches zero at high redshift, the relative Poisson uncertainty increases, limiting the constraining power of this indicator for testing numerical models.


In terms of absolute number density, most galaxy formation models systematically underestimate the abundance of high-mass QGs compared to our observations, though a few remain consistent within uncertainties at $z \lesssim 5$. The models achieving the best agreement at high redshift are generally those incorporating stronger AGN feedback mechanisms \citep[e.g., \texttt{GAEA} \& \texttt{SHARK};][]{fontanot_rise_2020,Lagos_diverse_2025}. This aligns with emerging observational evidence linking the rapid quenching of massive galaxies to supermassive black hole activity \citep[e.g.,][]{xie_first_2024,Onoue_pathway_2025,Wu_obermassive_2025}. A quantitative comparison with quasar demographics supports this interpretation. The number density of visible quasars at $z \sim 6$ is approximately $10^{-7}~\mathrm{Mpc^{-3}}$ \citep{McGreer_quasar_2013}, which is nominally much lower than the QG density. However, when accounting for the short duty cycle of the quasar phase, the total number density of supermassive black holes capable of driving feedback (including potential quasars) becomes comparable to that of the massive QG population we observe.

For low-mass galaxies, most models successfully reproduce the declining number density trend up to $z=6$. Beyond this epoch, however, \review2{most predictions show notable differences from observations}: models fail to replicate the observed ``plateau" in number density, dropping well below the measured values. The \texttt{Shark} (2024) model is an exception, successfully predicting the high abundance of low-mass QGs at $z > 6$, but at the cost of significantly overestimating them at $z < 6$. Interestingly, when the model was refined to adjust the RPS effect (in \texttt{Shark} 2026), the low-redshift overestimation was resolved, but the model returned to underestimating the $z > 6$ population, similar to other simulations. This ``see-saw" behavior highlights the complexity of modeling environmental quenching: mechanisms tuned to match the local universe may not capture the distinct environmental conditions of the early universe. This suggests that the interplay between environmental effects, feedback, and the halo assembly cycle at $z > 6$ requires further theoretical refinement.



However, we acknowledge that methodological differences may contribute to observation-model discrepancies. A key factor is the timescale used to define star formation rates. Galaxy formation models often average SFR over relatively long timescales (e.g., $\sim 200~\mathrm{Myr}$ for \texttt{GAEA} at $z > 5$). This averaging can smooth out ``bursty" histories, potentially leading to the omission of recently quenched PSBs that are identified as quiescent in observations. Future work requires a more direct comparison of the SFH between observed candidates and simulated galaxies to resolve this potential bias.

\subsection{Mass-dependent Evolutionary Histories of quiescent Galaxies at \texorpdfstring{$z \gtrsim 2$}{z > 4}}


For high-mass galaxies, we observe a steep decline in number density with increasing redshift. By $z \sim 8$, this population is nearly absent ($n < 10^{-6}~\mathrm{Mpc^{-3}}$) and constitutes a negligible fraction of the total population ($f_q \lesssim 2\%$ for QGs with $\log M_\star/M_\odot>9$). This scarcity implies that permanently quenching a massive galaxy at these early epochs is an extremely inefficient or rare process. Consequently, the extremely early-quenching massive galaxy reported by \cite{glazebrook_massive_2024} likely represents an exceptional outlier rather than a common population, consistent with the standard predictions of the $\Lambda$CDM framework.


\blue{In contrast to the high-mass regime, the number density of low-mass galaxies exhibits a remarkably flat evolution (a ``plateau") at $z > 5$. This divergence suggests that QGs follow fundamentally distinct evolutionary pathways depending on their mass. We propose that this plateau is a signature of temporary quiescence. Several theoretical studies suggest that high-redshift, low-mass QGs are not permanently ``dead" but are in a transient phase, likely to rejuvenate upon re-accreting gas \citep[e.g.,][]{Gelli_temporarilyquench_2025,Szpila_QG_2025}. In this scenario, galaxies constantly enter and leave the quiescent phase, maintaining a relatively stable number density—the observed plateau—rather than accumulating monotonically as seen in the permanent quenching model. Examining the SMF further supports this mass-dependent evolution. If we divide the sample at $\log M_{\star}/M_{\odot} \approx 10.0-10.3$, the low-mass population exhibits another plateau-like evolution at $2 < z < 4$, distinct from the rapid buildup of high-mass QGs. The significantly slower net growth rate of low-mass QGs during this period implies that a large fraction of them may be returning to the star-forming main sequence, effectively countering the production of newly quenched systems. A similar trend at $z \sim 3-5$ has also been noted by \cite{merlin_downsizing_2025}.}

\blue{While galaxy rejuvenation is considered rare at low redshift ($\lesssim 10\%$; \citealt{tacchella_fast_2022,tanaka_hinotori_2023}), simulations suggest its frequency increases to $\gtrsim 20\%$ at earlier times \citep{remus_relight_2023}. Our results point to an even more dynamic picture, suggesting that rejuvenation could be a dominant mechanism regulating the low-mass QG population at $z > 2$. Definitive confirmation of these bursty star-formation histories will require deep spectroscopic follow-up to constrain stellar ages and gas kinematics.}


\blue{The temporary nature of high-redshift low-mass QGs aligns with our understanding of quenching physics. Theoretical models and local observations indicate that fully quenching a central galaxy typically requires a massive black hole ($M_{\rm BH} \gtrsim 10^{7.5}M_{\odot}$) to maintain feedback and prevent cooling flows \citep{Bluck_BH_2023,Wang_BH_2024}. Low-mass galaxies at high redshift likely lack such supermassive black holes, making it difficult to achieve permanent quiescence (unless they are satellites). In contrast, the rapid rise in the number density of low-mass QGs at $z \lesssim 2$ is likely driven by the onset of environmental quenching as clusters and groups assemble \citep{peng_environment_2010}. This dichotomy in mechanisms is further supported by structural analysis: in a forthcoming paper (Chen et al. submitted; see also \citealt{Sato_IMQG_2024}), we observe an inflection point in the mass–size relation at $\log(M_\star/M_\odot) \approx 10.3$, suggesting that high- and low-mass QGs do not lie on the same sequence. This divergence in structural properties implies that massive galaxies are quenched via internal AGN feedback (leading to compact remnants), while low-mass galaxies are governed by environmental effects or feedback-driven temporal pauses.}

\subsection{Impact of MIRI Photometry and Mass Completeness}

\label{subsec:impact of MIRI}

\blue{To quantify the impact of MIRI photometry on QG selection, we isolate its influence by fixing redshifts to the MIRI-inclusive catalog values, following findings that MIRI adds minimal constraints to $z_{\mathrm{phot}}$ when high-quality HST/NIRCam data exist \citep{stevenson_primer_2025}. At $z \sim 3–5$, MIRI data primarily aid in the removal of dusty contaminants, reducing derived number densities by $\sim 0.1–0.2~\mathrm{dex}$—a systematic offset comparable to the statistical uncertainty at these redshifts. Crucially, the role of MIRI shifts at higher redshifts: at $z > 5$, it becomes {instrumental} for confirmation rather than rejection, with \review2{3/4} of our candidates with $\log M_\star/M_\odot>9.5$ requiring MIRI data for identification. This heavy reliance necessitates a careful examination of how MIRI's depth and completeness impact our high-redshift QG census.}

{To assess our selection completeness, we calibrated our detection limits against the deeper JADES fields. For galaxies with $\log M_{\star}/M_{\odot} > 9.5$ at $z \sim 8$, the NIRCam F444W mass completeness is estimated to exceed 95\%. Assuming the F444W selection is effectively complete, we derived the F770W completeness by calculating the recovery fraction of F444W sources in the F770W band.}

{For QGs, detection completeness depends on two competing factors: their higher mass-to-light (M/L) ratios (which reduce total flux) and their compact morphologies (which enhance surface brightness). In the F444W band, these two effects largely offset each other, rendering the completeness for QGs comparable to that of SFGs. In the F770W band, however, detections are limited by the coarser PSF of MIRI. At this resolution, the morphological compactness provides no advantage as both QGs and SFGs appear similarly unresolved. Consequently, the M/L effect dominates, making QGs fainter than SFGs at fixed stellar mass. We quantified this difference and found that at a fixed magnitude in F770W, QGs are on average 0.27 dex more massive than SFGs. Accounting for this M/L-driven mass shift, we estimate the mean F770W completeness for QGs with $9.5<\log M_{\star}/M_{\odot}<10.6$ at $z \sim 7$ to be approximately 76\%, and for the lowest mass bin $9.5<\log M_{\star}/M_{\odot}<10.2$ of SMF at \review2{$5<z<8$}, \review2{the completeness is around $70\%$}.}

 For sources falling below the $2\sigma$ detection threshold in F770W, we employ $3\sigma$ upper limits to constrain the SEDs. In our final sample, only one QG ($z_{\mathrm{phot}}=6.02$) and one mini-quench galaxy ($z_{\mathrm{phot}}=4.59$) rely explicitly on MIRI upper limits for their classification. Specifically, while the QG candidate remains robust even without MIRI constraints, the mini-quench candidate would be misclassified as a star-forming galaxy without the upper-limit constraints. This suggests that while direct MIRI detections are crucial, upper limits also play a valuable role in expanding the effective sample size beyond the nominal detection completeness.

{Since a subset of QGs can be identified even without direct MIRI detections, the actual completeness is likely higher than what implies by the MIRI depth alone. Consequently, assuming the undetected population mirrors the detected fraction, we estimate that MIRI depth limitations could lead to an underestimation of at most a factor of 1.3 ($\approx0.11~\mathrm{dex}$) for the low-mass QG number density at $6<z<8$, and a factor of 2 ($\approx0.3~\mathrm{dex}$) for the lowest mass bin of the SMF at \review2{$5<z<8$}. This maximum potential biases remain comparable to or smaller than the uncertainty for the number density measurements.} 

Critically, applying such a correction would only reinforce our main findings: it would make the observed number density plateau at low mass and high redshift even more pronounced, further widening the discrepancy with theoretical models. Given the limited sample size, we refrain from applying this uncertain correction and present our results as robust lower limits. To definitively characterize this low-mass population in the Epoch of Reionization, deeper MIRI observations will be indispensable.

\section{Summary}
\label{sec:summary}


This study presents a comprehensive analysis of the number density evolution of QGs up to $z=8$. We leverage the deep JWST/NIRCam and MIRI imaging from the PRIMER survey, covering $227~\mathrm{arcmin}^2$, combined with complementary space- and ground-based legacy data. Following a homogeneous reduction of JWST imaging, we derive photometric redshifts and physical properties using \texttt{EAZY} and \texttt{BAGPIPES}, respectively. Our selection strategy utilizes an sSFR threshold and incorporates a hybrid technique based on \cite{labbe_uncover_2023}, \cite{Rinaldi_dot_2025} and \cite{kocevski_rise_2024} to robustly exclude LRD contaminants. This yields a final sample of \review2{45} quiescent galaxy candidates at $3 < z < 8$. Our main findings are summarized as follows:

\begin{itemize}

 \item \textbf{MIRI is instrumental for QG purity and completeness at $z > 3$:} \blue{Mid-infrared coverage is critical for constraining the sSFR and stellar mass of high-redshift candidates. Crucially, MIRI plays a dual role: it helps to remove dusty star-forming contaminants at $z\sim 3-5$, and to recover low-mass QGs at $z \gtrsim 5$ that are often missed by using near-infrared data alone.}
 
 \item \textbf{High-mass QGs show rapid decline and potential AGN connection:} The number density of massive QGs ($\log (M_{\star}/M_{\odot}) > 10.6$) drops precipitously with redshift, from $n = 1.32\times10^{-5}~~\mathrm{Mpc^{-3}}$ at $z \sim 3-4$ to $n < 10^{-6}~~\mathrm{Mpc^{-3}}$ at $z > 6$. This scarcity at early epochs is quantitatively consistent with the space density of SDSS quasars ($n\sim10^{-7}~\mathrm{Mpc}^{-3}$) once duty cycles are accounted for, reinforcing the evolutionary link between luminous quasars and the formation of the first massive quiescent galaxies.
 
 \item \textbf{Low-mass QGs exhibit a ``Plateau" indicative of temporary quenching:} In contrast to their massive counterparts, low-mass QGs ($9.5 < \log (M_{\star}/M_{\odot}) < 10.6$) maintain a remarkably constant number density {($n \sim 2\times10^{-6}~\mathrm{Mpc^{-3}}$)} from $z \sim 4$ to $z \sim 8$. A similar plateau is observed for even lower-mass galaxies ($\log M_{\star}/M_{\odot} < 10.3$) at $z \sim 2-4$. We propose that this flat evolution signifies a dynamic equilibrium driven by \textit{temporary quenching}, where rejuvenation episodes are common among low-mass galaxies in the early Universe.
 
 \item \textbf{Tension with Galaxy Formation Models:} Most current theoretical models systematically underestimate the abundance of high-redshift QGs. The variance between models suggests that AGN feedback implementations are a key factor in reproducing the high-mass population. Furthermore, the observed ``plateau" in the low-mass regime is largely absent in simulations, indicating that the interplay of environmental effects, feedback, and gas replenishment cycles at high redshift requires significant theoretical refinement.
 \end{itemize}

While this work provides new constraints on the early quiescent population, we acknowledge certain limitations that define the path for future investigation. First, despite the depth of PRIMER, the sample size of QGs at $z > 4$ remains small (N $\sim 5$ per bin), resulting in substantial Poisson uncertainties. Second, approximately 70\% of our robust sample currently lacks spectroscopic confirmation. 
To achieve tighter statistical constraints on the high-redshift QG population, larger surveys with deep NIRCam and MIRI synergy are essential. This strategy does not only expand the sample size but also provides the robust, complete targets required for efficient spectroscopic follow-up, ultimately enabling us to map the full evolutionary history of these early galaxies.


\begin{acknowledgments}
This work was supported by National Natural Science Foundation of China (Grant No.12525302, 12173017 and 12141301), Natural Science Foundation of Jiangsu Higher Education Institutions of China(Grant No. BK20250001), National Key R\&D Program of China (Grant no. 2023YFA1605600), Scientific Research Innovation Capability Support Project for Young Faculty (Project No. ZYGXQNJSKYCXNLZCXM-P3), the Fundamental Research Funds for the Central Universities with Grant no.KG202502, and the China Manned Space Program with grant no. CMS-CSST-2025-A04. KW acknowledges support from the Science and Technologies Facilities Council (STFC) through grant ST/X001075/1. And S.L. acknowledges the support from the Key Laboratory of Modern Astronomy and Astrophysics (Nanjing University) by the Ministry of Education, whose work was supported by the National Natural Science Foundation of China (Project No. 12503011). 
Some of the data products presented herein were retrieved from the Dawn JWST Archive (DJA). DJA is an initiative of the Cosmic Dawn Center (DAWN), which is funded by the Danish National Research Foundation under grant DNRF140. The specific data from MAST analyzed can be accessed via doi:\href{https://archive.stsci.edu/doi/resolve/resolve.html?doi=10.17909/xcqt-gw25}{10.17909/xcqt-gw25}
\end{acknowledgments}

%

\vspace{5mm}


\software{EAZY \citep{brammer_eazy_2008},  
          BAGPIPES \citep{carnall_inferring_2018}          
          }


\clearpage
\appendix
\FloatBarrier
\setcounter{figure}{0}
\renewcommand{\thefigure}{A.\arabic{figure}}
\renewcommand{\thetable}{A.\arabic{table}}

\begin{deluxetable}{cccc}
    \label{table:bestfit}
    \tablecaption{Best-fit Parameters of the SMF of QGs}
    \tablehead{
        \colhead{Redshift} & \colhead{$M_* (\log M_\odot)$} & \colhead{$\phi^* (\times 10^{-5} \mathrm{Mpc^{-3}})$} & \colhead{$\alpha$}
    }
    \startdata
    2.0--2.5 & $10.40 \pm 0.04$ & $93.56 \pm 3.48$  & $0.84 \pm 0.17$ \\
    2.5--3.0 & $10.57 \pm 0.06$ & $52.03 \pm 1.77$  & $0.52 \pm 0.20$ \\
    3.0--3.5 & $10.35 \pm 0.09$ & $17.27 \pm 1.13$  & $0.09 \pm 0.26$ \\
    3.5--4.0 & $10.40 \pm 0.10$ & $16.45 \pm 1.10$  & $0.15 \pm 0.30$ \\
    4.0--5.0 & $10.35 \pm 0.09$ & $10.17 \pm 0.71$  & $0.56 \pm 0.22$ \\
    5.0--8.0 & $10.39 \pm 0.23$ & $4.35 \pm 0.69$   & $-0.30 \pm 0.58$ \\
    \enddata
    \tablecomments{Best-fit results for quiescent galaxies fitted with a single Schechter function \citep[see][for methodology]{sun_protocluster_2024}. Fitting results are shown in Figure \ref{fig:highz_SMF}.}
\end{deluxetable}

\bibliography{main}{}
\bibliographystyle{aasjournalv7}



\end{document}